\documentclass{article}
\usepackage{url}

\makeatletter
\@ifundefined{lhs2tex.lhs2tex.sty.read}%
  {\@namedef{lhs2tex.lhs2tex.sty.read}{}%
   \newcommand\SkipToFmtEnd{}%
   \newcommand\EndFmtInput{}%
   \long\def\SkipToFmtEnd#1\EndFmtInput{}%
  }\SkipToFmtEnd

\newcommand\ReadOnlyOnce[1]{\@ifundefined{#1}{\@namedef{#1}{}}\SkipToFmtEnd}
\usepackage{amstext}
\usepackage{amssymb}
\usepackage{stmaryrd}
\DeclareFontFamily{OT1}{cmtex}{}
\DeclareFontShape{OT1}{cmtex}{m}{n}
  {<5><6><7><8>cmtex8
   <9>cmtex9
   <10><10.95><12><14.4><17.28><20.74><24.88>cmtex10}{}
\DeclareFontShape{OT1}{cmtex}{m}{it}
  {<-> ssub * cmtt/m/it}{}

\DeclareFontShape{OT1}{cmtt}{bx}{n}
  {<5><6><7><8>cmtt8
   <9>cmbtt9
   <10><10.95><12><14.4><17.28><20.74><24.88>cmbtt10}{}
\DeclareFontShape{OT1}{cmtex}{bx}{n}
  {<-> ssub * cmtt/bx/n}{}

\newcommand{\Conid}[1]{\mathit{#1}}
\newcommand{\Varid}[1]{\mathit{#1}}
\newcommand{\anonymous}{\kern0.06em \vbox{\hrule\@width.5em}}

\renewcommand{\leq}{\leqslant}
\renewcommand{\geq}{\geqslant}
\usepackage{polytable}

\@ifundefined{mathindent}%
  {\newdimen\mathindent\mathindent\leftmargini}%
  {}%

\def\resethooks{%
  \global\let\SaveRestoreHook\empty
  \global\let\ColumnHook\empty}
\newcommand*{\savecolumns}[1][default]%
  {\g@addto@macro\SaveRestoreHook{\savecolumns[#1]}}
\newcommand*{\restorecolumns}[1][default]%
  {\g@addto@macro\SaveRestoreHook{\restorecolumns[#1]}}
\newcommand*{\aligncolumn}[2]%
  {\g@addto@macro\ColumnHook{\column{#1}{#2}}}

\resethooks

\newcommand{\onelinecommentchars}{\quad-{}- }
\newcommand{\commentbeginchars}{\enskip\{-}
\newcommand{\commentendchars}{-\}\enskip}

\newcommand{\visiblecomments}{%
  \let\onelinecomment=\onelinecommentchars
  \let\commentbegin=\commentbeginchars
  \let\commentend=\commentendchars}

\newcommand{\invisiblecomments}{%
  \let\onelinecomment=\empty
  \let\commentbegin=\empty
  \let\commentend=\empty}

\visiblecomments

\newlength{\blanklineskip}
\newcommand{\hsindent}[1]{\quad}
\let\hspre\empty
\let\hspost\empty

\EndFmtInput
\makeatother
\ReadOnlyOnce{polycode.fmt}%
\makeatletter

\newcommand{\hsnewpar}[1]%
  {{\parskip=0pt\parindent=0pt\par\vskip #1\noindent}}

\newcommand{\hscodestyle}{}

\newcommand{\sethscode}[1]%
  {\expandafter\let\expandafter\hscode\csname #1\endcsname
   \expandafter\let\expandafter\endhscode\csname end#1\endcsname}

  {\par\noindent
   \advance\leftskip\mathindent
   \hscodestyle
   \let\\=\@normalcr
   \let\hspre\(\let\hspost\)%
   \pboxed}%
  {\endpboxed\)%
   \par\noindent
   \ignorespacesafterend}

  {\hsnewpar\abovedisplayskip
   \advance\leftskip\mathindent
   \hscodestyle
   \let\hspre\(\let\hspost\)%
   \pboxed}%
  {\endpboxed%
   \hsnewpar\belowdisplayskip
   \ignorespacesafterend}

  {\hsnewpar\abovedisplayskip
   \advance\leftskip\mathindent
   \hscodestyle
   \let\\=\@normalcr
   \(\pboxed}%
  {\endpboxed\)%
   \hsnewpar\belowdisplayskip
   \ignorespacesafterend}

\newcommand{\plainhs}{\sethscode{plainhscode}}

\plainhs

  {\hsnewpar\abovedisplayskip
   \advance\leftskip\mathindent
   \hscodestyle
   \let\\=\@normalcr
   \(\parray}%
  {\endparray\)%
   \hsnewpar\belowdisplayskip
   \ignorespacesafterend}

  {\parray}{\endparray}

  {\(\parray}{\endparray\)}

\def\codeframewidth{\arrayrulewidth}
\RequirePackage{calc}

  {\parskip=\abovedisplayskip\par\noindent
   \hscodestyle
   \arrayrulewidth=\codeframewidth
   \tabular{@{}|p{\linewidth-2\arraycolsep-2\arrayrulewidth-2pt}|@{}}%
   \hline\framedhslinecorrect\\{-1.5ex}%
   \let\endoflinesave=\\
   \let\\=\@normalcr
   \(\pboxed}%
  {\endpboxed\)%
   \framedhslinecorrect\endoflinesave{.5ex}\hline
   \endtabular
   \parskip=\belowdisplayskip\par\noindent
   \ignorespacesafterend}

\newcommand{\framedhslinecorrect}[2]%
  {#1[#2]}

  {\(\def\column##1##2{}%
   \let\>\undefined\let\<\undefined\let\\\undefined
   \newcommand\>[1][]{}\newcommand\<[1][]{}\newcommand\\[1][]{}%
   \def\fromto##1##2##3{##3}%
   }{\) }%

  {\let\orighscode=\hscode
   \let\origendhscode=\endhscode
   \def\endhscode{\def\hscode{\endgroup\def\@currenvir{hscode}\\}\begingroup}
   \orighscode\def\hscode{\endgroup\def\@currenvir{hscode}}}%
  {\origendhscode
   \global\let\hscode=\orighscode
   \global\let\endhscode=\origendhscode}%

\makeatother
\EndFmtInput
\ReadOnlyOnce{forall.fmt}%
\makeatletter

\let\HaskellResetHook\empty
\newcommand*{\AtHaskellReset}[1]{%
  \g@addto@macro\HaskellResetHook{#1}}
\newcommand*{\HaskellReset}{\HaskellResetHook}

\newcommand\hsforall{\global\let\hsdot=\hsperiodonce}
\newcommand*\hsperiodonce[2]{#2\global\let\hsdot=\hscompose}
\newcommand*\hscompose[2]{#1}

\AtHaskellReset{\global\let\hsdot=\hscompose}

\HaskellReset

\makeatother
\EndFmtInput
\ReadOnlyOnce{exists.fmt}%
\makeatletter

\newcommand\hsexists{\global\let\hsdot=\hsperiodonce}

\AtHaskellReset{\global\let\hsdot=\hscompose}

\HaskellReset

\makeatother
\EndFmtInput

\usepackage{classicthesis}
\usepackage{amsmath}
\usepackage{natbib}
\usepackage{doubleequals}

\def\commentbegin{\quad\begingroup\color{Green}\{\ }
\def\commentend{\}\endgroup}
\mathindent=\parindent

\allowdisplaybreaks

\definecolor{mediumpersianblue}{rgb}{0.0, 0.4, 0.65}
\everymath{\color{mediumpersianblue}}

\begin{document}

\title{{\large\bf Functional Pearl}\\
A Greedy Algorithm for Dropping Digits}

\author{\color{black}Richard Bird$^1$ \and \color{black}Shin-Cheng Mu$^2$}
\date{%
    $^1$Department of Computer Science, University of Oxford\\%
    $^2$Institute of Information Science, Academia Sinica
}

\maketitle

\begin{abstract}
Consider the puzzle: given a number, remove \ensuremath{\Varid{k}} digits such that the resulting number is as large as possible.
Various techniques were employed to derive a linear-time solution to the puzzle:
predicate logic was used to justify the structure of a greedy algorithm, a dependently-typed proof assistant was used to give a constructive proof of the greedy condition, and equational reasoning was used to calculate the greedy step as well as the final, linear-time optimisation.
\end{abstract}

\section{Introduction}

Greedy algorithms abound in computing.
Well-known examples include Huffman coding, minimum cost spanning trees, and the coin-changing problem; but there are many others.
This pearl adds yet another problem to this collection.
However, as has been said before, greedy algorithms can be tricky things.
The trickiness is not in the algorithm itself, which is usually quite short and easy to understand, but in the proof that it does produce a best possible result.
The mathematical theory of \emph{matroids}, see~\cite{Lawler:76:Combinatorial}, and its generalisation to \emph{greedoids}, see~\cite{Korte:91:Greedoids}, have been developed to explain why and when many greedy algorithms work, although the theory does not cover all possible cases.
In practice, greedy algorithms are usually verified directly rather than by extracting the underlying matroid or greedoid. \cite{Curtis:03:Classification} discusses four basic ways in which a greedy algorithm can be proved to work, one of which will be followed with our problem.

The problem is to remove \ensuremath{\Varid{k}} digits from a number containing at least \ensuremath{\Varid{k}} digits, so that the result is as large as possible.
For example, removing one digit from the number \ensuremath{\text{\ttfamily \char34 6782334\char34}} gives \ensuremath{\text{\ttfamily \char34 782334\char34}} as the largest possible result, while removing three digit yields \ensuremath{\text{\ttfamily \char34 8334\char34}}.
Given that a number can be seen as a list of digits,
the problem can be generalised to removing, from a list whose elements are drawn from a linearly ordered type, \ensuremath{\Varid{k}} elements so that the result is largest under lexicographic ordering.
While the problem was invented out of curiosity rather than for a pressing application, it has apparently been used as an interview question for candidates seeking jobs in computing.%
\footnote{The problem is listed on LeetCode as \url{https://leetcode.com/problems/remove-k-digits/}, where the objective is to find the smallest number rather than the largest, but the principles
are the same.} The hope is that we can discover an algorithm that takes linear time in the number of elements.

The first task is to give a formal specification of the problem.
Consider the function \ensuremath{\Varid{drops}} that removes a single element from
a non-empty list in all possible ways:
\footnote{We use notations similar to Haskell, with slight variations.
For example, the type for lists is denoted by \ensuremath{\Conid{List}}, and we allow \ensuremath{(\mathrm{1}\mathbin{+}\Varid{k})} as a pattern in function definitions, to match our inductive proofs.
Laziness is not needed.
}
\begin{hscode}\SaveRestoreHook
\column{B}{@{}>{\hspre}l<{\hspost}@{}}%
\column{15}{@{}>{\hspre}c<{\hspost}@{}}%
\column{15E}{@{}l@{}}%
\column{18}{@{}>{\hspre}l<{\hspost}@{}}%
\column{E}{@{}>{\hspre}l<{\hspost}@{}}%
\>[B]{}\Varid{drops}\mathbin{::}\Conid{List}\;\Varid{a}\to \Conid{List}\;(\Conid{List}\;\Varid{a}){}\<[E]%
\\
\>[B]{}\Varid{drops}\;[\mskip1.5mu \Varid{x}\mskip1.5mu]{}\<[15]%
\>[15]{}\mathrel{=}{}\<[15E]%
\>[18]{}[\mskip1.5mu [\mskip1.5mu \mskip1.5mu]\mskip1.5mu]{}\<[E]%
\\
\>[B]{}\Varid{drops}\;(\Varid{x}\mathbin{:}\Varid{xs}){}\<[15]%
\>[15]{}\mathrel{=}{}\<[15E]%
\>[18]{}\Varid{xs}\mathbin{:}\Varid{map}\;(\Varid{x}\mathbin{:})\;(\Varid{drops}\;\Varid{xs}){}\<[E]%
\ColumnHook
\end{hscode}\resethooks
For example, \ensuremath{\Varid{drops}\;\text{\ttfamily \char34 abcd\char34}\mathrel{=}[\mskip1.5mu \text{\ttfamily \char34 bcd\char34},\text{\ttfamily \char34 acd\char34},\text{\ttfamily \char34 abd\char34},\text{\ttfamily \char34 abc\char34}\mskip1.5mu]}.
The function \ensuremath{\Varid{solve}} for computing a solution can be defined by a simple
exhaustive search:
\begin{hscode}\SaveRestoreHook
\column{B}{@{}>{\hspre}l<{\hspost}@{}}%
\column{10}{@{}>{\hspre}c<{\hspost}@{}}%
\column{10E}{@{}l@{}}%
\column{13}{@{}>{\hspre}l<{\hspost}@{}}%
\column{E}{@{}>{\hspre}l<{\hspost}@{}}%
\>[B]{}\Varid{solve}\mathbin{::}\Conid{Ord}\;\Varid{a}\Rightarrow \Conid{Nat}\to \Conid{List}\;\Varid{a}\to \Conid{List}\;\Varid{a}{}\<[E]%
\\
\>[B]{}\Varid{solve}\;\Varid{k}{}\<[10]%
\>[10]{}\mathrel{=}{}\<[10E]%
\>[13]{}\Varid{maximum}\hsdot{\cdot }{.}\Varid{apply}\;\Varid{k}\;\Varid{step}\hsdot{\cdot }{.}\Varid{wrap}~~,{}\<[E]%
\\[\blanklineskip]%
\>[B]{}\Varid{step}\mathbin{::}\Conid{List}\;(\Conid{List}\;\Varid{a})\to \Conid{List}\;(\Conid{List}\;\Varid{a}){}\<[E]%
\\
\>[B]{}\Varid{step}\mathrel{=}\Varid{concat}\hsdot{\cdot }{.}\Varid{map}\;\Varid{drops}~~.{}\<[E]%
\ColumnHook
\end{hscode}\resethooks
The function \ensuremath{\Varid{solve}} converts a given input into a singleton list,
applies the function \ensuremath{\Varid{step}} exactly \ensuremath{\Varid{k}} times to produce all possible candidates, and computes the lexical maximum of the result.
\ensuremath{\Conid{Nat}} is the type of natural numbers.
The function \ensuremath{\Varid{step}} is \ensuremath{\Varid{drops}} lifted to a list of candidates.
It computes, for each candidate, all the ways to drop \ensuremath{\mathrm{1}} element.
Functions \ensuremath{\Varid{wrap}} and \ensuremath{\Varid{apply}} are respectively defined by\\
\hspace{2cm}
\begin{minipage}{0.3\textwidth}
\begin{hscode}\SaveRestoreHook
\column{B}{@{}>{\hspre}l<{\hspost}@{}}%
\column{E}{@{}>{\hspre}l<{\hspost}@{}}%
\>[B]{}\Varid{wrap}\mathbin{::}\Varid{a}\to \Conid{List}\;\Varid{a}{}\<[E]%
\\
\>[B]{}\Varid{wrap}\;\Varid{x}\mathrel{=}[\mskip1.5mu \Varid{x}\mskip1.5mu]~~,{}\<[E]%
\ColumnHook
\end{hscode}\resethooks
\end{minipage}
\begin{minipage}{0.4\textwidth}
\begin{hscode}\SaveRestoreHook
\column{B}{@{}>{\hspre}l<{\hspost}@{}}%
\column{14}{@{}>{\hspre}l<{\hspost}@{}}%
\column{E}{@{}>{\hspre}l<{\hspost}@{}}%
\>[B]{}\Varid{apply}\mathbin{::}\Conid{Nat}\to (\Varid{a}\to \Varid{a})\to \Varid{a}\to \Varid{a}{}\<[E]%
\\
\>[B]{}\Varid{apply}\;\mathrm{0}\;{}\<[14]%
\>[14]{}\Varid{f}\mathrel{=}\Varid{id}{}\<[E]%
\\
\>[B]{}\Varid{apply}\;(\mathrm{1}\mathbin{+}\Varid{k})\;{}\<[14]%
\>[14]{}\Varid{f}\mathrel{=}\Varid{apply}\;\Varid{k}\;\Varid{f}\hsdot{\cdot }{.}\Varid{f}~~,{}\<[E]%
\ColumnHook
\end{hscode}\resethooks
\end{minipage}\\
For brevity, for the rest of the pearl we will write \ensuremath{\Varid{apply}\;\Varid{k}\;\Varid{f}} as \ensuremath{{\Varid{f}}^{\Varid{k}}}.
Since a sequence of length \ensuremath{\Varid{n}} has \ensuremath{\Varid{n}} drops, and computing the larger of two lists of
length \ensuremath{\Varid{n}\mathbin{-}\Varid{k}} takes \ensuremath{\Conid{O}\;(\Varid{n}\mathbin{-}\Varid{k})} steps, this method for computing the answer
takes \ensuremath{\Conid{O}\;(\Varid{n}^{\Varid{k}})} steps. We aim to do better.

\section{A Greedy Algorithm}

To obtain a greedy algorithm, one would wish that the best way to remove \ensuremath{\Varid{k}} digits can be computed by removing \ensuremath{\mathrm{1}} digit \ensuremath{\Varid{k}} times, and each time we greedily remove the digit that makes the current result as large as possible.
That is, letting
\begin{hscode}\SaveRestoreHook
\column{B}{@{}>{\hspre}l<{\hspost}@{}}%
\column{E}{@{}>{\hspre}l<{\hspost}@{}}%
\>[B]{}\Varid{gstep}\mathbin{::}\Conid{Ord}\;\Varid{a}\Rightarrow \Conid{List}\;\Varid{a}\to \Conid{List}\;\Varid{a}{}\<[E]%
\\
\>[B]{}\Varid{gstep}\mathrel{=}\Varid{maximum}\hsdot{\cdot }{.}\Varid{drops}~~,{}\<[E]%
\ColumnHook
\end{hscode}\resethooks
we wish to have
\begin{align}
\ensuremath{\Varid{maximum}\hsdot{\cdot }{.}{\Varid{step}}^{\Varid{k}}\hsdot{\cdot }{.}\Varid{wrap}\mathrel{=}{\Varid{gstep}}^{\Varid{k}}~~.} \label{eq:k-gsteps-correct}
\end{align}

One cannot just claim that this strategy works without proper reasoning, however.
It can be shown that \eqref{eq:k-gsteps-correct} is true if the following monotonicity condition holds
(we denote lexicographic ordering on lists by \ensuremath{(\unlhd)}, and ordering on individual elements by \ensuremath{(\leq )}):
\begin{align}
  \ensuremath{\Varid{xs}\unlhd\Varid{ys}~\Rightarrow~(\forall \Varid{xs'}\hsforall \in\Varid{drops}\;\Varid{xs}\mathbin{:}(\exists \Varid{ys'}\hsexists \in\Varid{drops}\;\Varid{ys}\mathbin{:}\Varid{xs'}\unlhd\Varid{ys'}))~~,}
  \label{eq:mono1}
\end{align}
where \ensuremath{(\in)} is overloaded to denote membership for lists.
That is, if \ensuremath{\Varid{ys}} is no worse than \ensuremath{\Varid{xs}}, whatever candidate we can obtain from \ensuremath{\Varid{xs}}, we can compute a candidate from \ensuremath{\Varid{ys}} that is no worse either.

Unfortunately, \eqref{eq:mono1} does not hold for our problem.
Consider \ensuremath{\Varid{xs}\mathrel{=}\text{\ttfamily \char34 1934\char34}\lhd\text{\ttfamily \char34 4234\char34}\mathrel{=}\Varid{ys}}, \ensuremath{\text{\ttfamily \char34 934\char34}} is a possible result of \ensuremath{\Varid{drops}\;\Varid{xs}}, but the best we can do by removing one digit from \ensuremath{\Varid{ys}} is \ensuremath{\text{\ttfamily \char34 434\char34}}.
Note that \eqref{eq:mono1} does not hold even if we restrict \ensuremath{\Varid{xs}} and \ensuremath{\Varid{ys}} to lists that can be obtained from the same source --- certainly both \ensuremath{\text{\ttfamily \char34 1934\char34}} and \ensuremath{\text{\ttfamily \char34 4234\char34}} are both result of removing two digits from, say, \ensuremath{\text{\ttfamily \char34 194234\char34}}.

In the terminology of \cite{Curtis:03:Classification}, the \emph{Better-Global} principle ---
which says that if one first step is no worse than another, then there is a global solution using the former that is no worse than one using the latter --- does not hold for this problem.
What does hold is a weaker property, the \emph{Best-Global} principle: a globally optimal solution can be obtained by starting out with the \emph{best} possible step.
Formally, what we do have is that for all \ensuremath{\Varid{xs}} and \ensuremath{\Varid{k}}:
\begin{equation}
\begin{split}
&\ensuremath{(\forall \Varid{xs'}\hsforall \in{\Varid{step}}^{\mathrm{1}\mathbin{+}\Varid{k}}\;[\mskip1.5mu \Varid{xs}\mskip1.5mu]\mathbin{:}} \\
& \qquad \ensuremath{(\exists \Varid{zs}\hsexists \in({\Varid{step}}^{\Varid{k}}\cdot\Varid{wrap}\cdot\Varid{gstep})\;\Varid{xs}\mathbin{:}\Varid{xs'}\unlhd\Varid{zs}))~~.}
\end{split}
\label{eq:step-gstep}
\end{equation}
That is, letting \ensuremath{\Varid{xs'}} be an arbitrary result of dropping \ensuremath{\Varid{k}\mathbin{+}\mathrm{1}} elements from \ensuremath{\Varid{xs}},
one can always obtain a result \ensuremath{\Varid{zs}} that is no worse than \ensuremath{\Varid{xs'}} by greedily dropping the best element (by \ensuremath{\Varid{gstep}}) and then dropping arbitrary \ensuremath{\Varid{k}} elements.

Property \eqref{eq:step-gstep} will be proved in Section~\ref{sec:greedy-condition}.
For now, let us see how \eqref{eq:step-gstep} helps to prove \eqref{eq:k-gsteps-correct}, that is,
\ensuremath{\Varid{maixmum}\cdot{\Varid{step}}^{\Varid{k}}\cdot\Varid{wrap}\mathrel{=}{\Varid{gstep}}^{\Varid{k}}}.
The proof proceeds by induction on \ensuremath{\Varid{k}}.
For \ensuremath{\Varid{k}\mathbin{:=}\mathrm{0}} both sides reduce to \ensuremath{\Varid{id}}.
For the inductive case we need the universal property of \ensuremath{\Varid{maximum}}:
for all \ensuremath{\Varid{s}\mathbin{::}\Varid{a}\to \Varid{b}} and \ensuremath{\Varid{p}\mathbin{::}\Varid{a}\to \Conid{List}\;\Varid{b}}, and for total order \ensuremath{(\unlhd)} on \ensuremath{\Varid{b}}:
\footnote{
We restrict our disucssion to total orders to ensure that \ensuremath{\Varid{maximum}} returns one unique result.
More general scenarios are discussed in \cite{BirddeMoor:97:Algebra}.
}
\begin{align*}
\begin{split}
\ensuremath{\Varid{s}\mathrel{=}\Varid{maximum}\cdot\Varid{p}~\mathrel{\equiv}~}&
        \ensuremath{(\forall \Varid{x}\hsforall \mathbin{:}\Varid{s}\;\Varid{x}\in\Varid{p}\;\Varid{x})\mathrel{\wedge}}              \\
       &\quad \ensuremath{(\forall \Varid{x}\hsforall ,\Varid{y}\mathbin{:}\Varid{y}\in\Varid{p}\;\Varid{x}\Rightarrow\Varid{y}\unlhd\Varid{s}\;\Varid{x})~~.}
\end{split}
\end{align*}
To prove \ensuremath{\Varid{maixmum}\cdot{\Varid{step}}^{\mathrm{1}\mathbin{+}\Varid{k}}\cdot\Varid{wrap}\mathrel{=}{\Varid{gstep}}^{\mathrm{1}\mathbin{+}\Varid{k}}} we need to show that
1. for all \ensuremath{\Varid{xs}}, \ensuremath{{\Varid{gstep}}^{\mathrm{1}\mathbin{+}\Varid{k}}\;\Varid{xs}} is a member of \ensuremath{{\Varid{step}}^{\mathrm{1}\mathbin{+}\Varid{k}}\;[\mskip1.5mu \Varid{xs}\mskip1.5mu]}, which is a routine proof, and
2. for all \ensuremath{\Varid{xs}} and for all \ensuremath{\Varid{ys}\in{\Varid{step}}^{\mathrm{1}\mathbin{+}\Varid{k}}\;[\mskip1.5mu \Varid{xs}\mskip1.5mu]}, we have
\ensuremath{\Varid{ys}\unlhd{\Varid{gstep}}^{\mathrm{1}\mathbin{+}\Varid{k}}\;\Varid{xs}}.
We reason:
\begin{hscode}\SaveRestoreHook
\column{B}{@{}>{\hspre}c<{\hspost}@{}}%
\column{BE}{@{}l@{}}%
\column{6}{@{}>{\hspre}l<{\hspost}@{}}%
\column{8}{@{}>{\hspre}l<{\hspost}@{}}%
\column{E}{@{}>{\hspre}l<{\hspost}@{}}%
\>[6]{}\Varid{ys}\unlhd{\Varid{gstep}}^{\mathrm{1}\mathbin{+}\Varid{k}}\;\Varid{xs}{}\<[E]%
\\
\>[B]{}\mathrel{\equiv}{}\<[BE]%
\>[6]{}\Varid{ys}\unlhd{\Varid{gstep}}^{\Varid{k}}\;(\Varid{gstep}\;\Varid{xs}){}\<[E]%
\\
\>[B]{}\mathrel{\equiv}{}\<[BE]%
\>[8]{}\mbox{\commentbegin  induction hypothesis  \commentend}{}\<[E]%
\\
\>[B]{}\hsindent{6}{}\<[6]%
\>[6]{}\Varid{ys}\unlhd\Varid{maximum}\;({\Varid{step}}^{\Varid{k}}\;[\mskip1.5mu \Varid{gstep}\;\Varid{xs}\mskip1.5mu]){}\<[E]%
\\
\>[B]{}\mathrel{\equiv}{}\<[BE]%
\>[8]{}\mbox{\commentbegin  since \ensuremath{\Varid{ys}\unlhd\Varid{maximum}\;\Varid{xss}~\mathrel{\equiv}~(\exists \Varid{zs}\hsexists \in\Varid{xss}\mathbin{:}\Varid{ys}\unlhd\Varid{zs})}  \commentend}{}\<[E]%
\\
\>[B]{}\hsindent{6}{}\<[6]%
\>[6]{}(\exists \Varid{xs}\hsexists \in{\Varid{step}}^{\Varid{k}}\;[\mskip1.5mu \Varid{gstep}\;\Varid{xs}\mskip1.5mu]\mathbin{:}\Varid{ys}\unlhd\Varid{xs}){}\<[E]%
\\
\>[B]{}\Leftarrow{}\<[BE]%
\>[8]{}\mbox{\commentbegin  by \eqref{eq:step-gstep}  \commentend}{}\<[E]%
\\
\>[B]{}\hsindent{6}{}\<[6]%
\>[6]{}\Varid{ys}\in{\Varid{step}}^{\mathrm{1}\mathbin{+}\Varid{k}}\;[\mskip1.5mu \Varid{xs}\mskip1.5mu]~~,{}\<[E]%
\ColumnHook
\end{hscode}\resethooks
which is our assumption. We have thus proved \eqref{eq:k-gsteps-correct}.

{\bf Remark}: the proof above was carried out using predicate logic.
There is a relational counterpart, in the style of \cite{BirddeMoor:97:Algebra}, that is slightly more concise and more general, which we unfortunately cannot present without adding a section introducing the notations and rules.

\section{Refining the Greedy Step}

We will prove the greedy condition \eqref{eq:step-gstep} in Section \ref{sec:greedy-condition}.
It will turn out that the proof makes use of properties of \ensuremath{\Varid{gstep}} that will be evident from its inductive definition.
Therefore we calculate an inductive definition of \ensuremath{\Varid{gstep}} in this section.

It is easy to derive \ensuremath{\Varid{gstep}\;[\mskip1.5mu \Varid{x}\mskip1.5mu]\mathrel{=}[\mskip1.5mu \mskip1.5mu]}. For the inductive case we reason
\begin{hscode}\SaveRestoreHook
\column{B}{@{}>{\hspre}c<{\hspost}@{}}%
\column{BE}{@{}l@{}}%
\column{7}{@{}>{\hspre}l<{\hspost}@{}}%
\column{8}{@{}>{\hspre}l<{\hspost}@{}}%
\column{9}{@{}>{\hspre}l<{\hspost}@{}}%
\column{E}{@{}>{\hspre}l<{\hspost}@{}}%
\>[7]{}\Varid{gstep}\;(\Varid{x}\mathbin{:}\Varid{y}\mathbin{:}\Varid{xs}){}\<[E]%
\\
\>[B]{}\mathrel{=}{}\<[BE]%
\>[8]{}\mbox{\commentbegin  definition of \ensuremath{\Varid{gstep}}  \commentend}{}\<[E]%
\\
\>[B]{}\hsindent{7}{}\<[7]%
\>[7]{}\Varid{maximum}\;(\Varid{drops}\;(\Varid{x}\mathbin{:}\Varid{y}\mathbin{:}\Varid{xs})){}\<[E]%
\\
\>[B]{}\mathrel{=}{}\<[BE]%
\>[8]{}\mbox{\commentbegin  definition of \ensuremath{\Varid{drops}}  \commentend}{}\<[E]%
\\
\>[B]{}\hsindent{7}{}\<[7]%
\>[7]{}\Varid{maximum}\;((\Varid{y}\mathbin{:}\Varid{xs})\mathbin{:}\Varid{map}\;(\Varid{x}\mathbin{:})\;(\Varid{drops}\;(\Varid{y}\mathbin{:}\Varid{xs}))){}\<[E]%
\\
\>[B]{}\mathrel{=}{}\<[BE]%
\>[8]{}\mbox{\commentbegin  definition of \ensuremath{\Varid{maximum}}  \commentend}{}\<[E]%
\\
\>[B]{}\hsindent{7}{}\<[7]%
\>[7]{}\Varid{max}\;(\Varid{y}\mathbin{:}\Varid{xs})\;(\Varid{maximum}\;(\Varid{map}\;(\Varid{x}\mathbin{:})\;(\Varid{drops}\;(\Varid{y}\mathbin{:}\Varid{xs})))){}\<[E]%
\\
\>[B]{}\mathrel{=}{}\<[BE]%
\>[8]{}\mbox{\commentbegin  since \ensuremath{\Varid{maximum}\;(\Varid{map}\;(\Varid{x}\mathbin{:})\;\Varid{xss})\mathrel{=}\Varid{x}\mathbin{:}\Varid{maximum}\;\Varid{xss}}, provided \ensuremath{\Varid{xss}} is nonempty  \commentend}{}\<[E]%
\\
\>[B]{}\hsindent{7}{}\<[7]%
\>[7]{}\Varid{max}\;(\Varid{y}\mathbin{:}\Varid{xs})\;(\Varid{x}\mathbin{:}\Varid{maximum}\;(\Varid{drops}\;(\Varid{y}\mathbin{:}\Varid{xs}))){}\<[E]%
\\
\>[B]{}\mathrel{=}{}\<[BE]%
\>[8]{}\mbox{\commentbegin  definition of \ensuremath{\Varid{gstep}}  \commentend}{}\<[E]%
\\
\>[B]{}\hsindent{7}{}\<[7]%
\>[7]{}\Varid{max}\;(\Varid{y}\mathbin{:}\Varid{xs})\;(\Varid{x}\mathbin{:}\Varid{gstep}\;(\Varid{y}\mathbin{:}\Varid{xs})){}\<[E]%
\\
\>[B]{}\mathrel{=}{}\<[BE]%
\>[8]{}\mbox{\commentbegin  definition of \ensuremath{\Varid{max}} and lexicographic ordering  \commentend}{}\<[E]%
\\
\>[B]{}\hsindent{7}{}\<[7]%
\>[7]{}\mathbf{if}\;\Varid{x}\mathbin{<}\Varid{y}\;\mathbf{then}\;\Varid{y}\mathbin{:}\Varid{xs}{}\<[E]%
\\
\>[7]{}\hsindent{1}{}\<[8]%
\>[8]{}\mathbf{else}\;\mathbf{if}\;\Varid{x}\doubleequals\Varid{y}\;\mathbf{then}\;\Varid{x}\mathbin{:}\Varid{max}\;\Varid{xs}\;(\Varid{gstep}\;(\Varid{y}\mathbin{:}\Varid{xs})){}\<[E]%
\\
\>[8]{}\hsindent{1}{}\<[9]%
\>[9]{}\mathbf{else}\;\Varid{x}\mathbin{:}\Varid{gstep}\;(\Varid{y}\mathbin{:}\Varid{xs}){}\<[E]%
\\
\>[B]{}\mathrel{=}{}\<[BE]%
\>[8]{}\mbox{\commentbegin  since \ensuremath{\Varid{xs}\unlhd\Varid{gstep}\;(\Varid{y}\mathbin{:}\Varid{xs})}  \commentend}{}\<[E]%
\\
\>[B]{}\hsindent{7}{}\<[7]%
\>[7]{}\mathbf{if}\;\Varid{x}\mathbin{<}\Varid{y}\;\mathbf{then}\;\Varid{y}\mathbin{:}\Varid{xs}\;\mathbf{else}\;\Varid{x}\mathbin{:}\Varid{gstep}\;(\Varid{y}\mathbin{:}\Varid{xs})~~.{}\<[E]%
\ColumnHook
\end{hscode}\resethooks
Hence we have
\begin{hscode}\SaveRestoreHook
\column{B}{@{}>{\hspre}l<{\hspost}@{}}%
\column{17}{@{}>{\hspre}c<{\hspost}@{}}%
\column{17E}{@{}l@{}}%
\column{20}{@{}>{\hspre}l<{\hspost}@{}}%
\column{E}{@{}>{\hspre}l<{\hspost}@{}}%
\>[B]{}\Varid{gstep}\;[\mskip1.5mu \Varid{x}\mskip1.5mu]{}\<[17]%
\>[17]{}\mathrel{=}{}\<[17E]%
\>[20]{}[\mskip1.5mu \mskip1.5mu]{}\<[E]%
\\
\>[B]{}\Varid{gstep}\;(\Varid{x}\mathbin{:}\Varid{y}\mathbin{:}\Varid{xs}){}\<[17]%
\>[17]{}\mathrel{=}{}\<[17E]%
\>[20]{}\mathbf{if}\;\Varid{x}\mathbin{<}\Varid{y}\;\mathbf{then}\;\Varid{y}\mathbin{:}\Varid{xs}\;\mathbf{else}\;\Varid{x}\mathbin{:}\Varid{gstep}\;(\Varid{y}\mathbin{:}\Varid{xs})~~.{}\<[E]%
\ColumnHook
\end{hscode}\resethooks
It turns out that \ensuremath{\Varid{gstep}\;\Varid{xs}} deletes \emph{the last element of
the longest descending prefix} of \ensuremath{\Varid{xs}}.
For easy reference, we will refer to this element as the \emph{hill foot} of the list.
For example, \ensuremath{\Varid{gstep}\;\text{\ttfamily \char34 8766678\char34}\mathrel{=}\text{\ttfamily \char34 876678\char34}}, where the hill foot, the element deleted, is the third \ensuremath{\mathrm{6}}.

\section{Proving the Greedy Condition}
\label{sec:greedy-condition}

In this section we aim to prove \eqref{eq:step-gstep}, recited here:
\begin{align*}
\ensuremath{(\forall \Varid{xs'}\hsforall \in{\Varid{step}}^{\mathrm{1}\mathbin{+}\Varid{k}}\;[\mskip1.5mu \Varid{xs}\mskip1.5mu]\mathbin{:}(\exists \Varid{zs}\hsexists \in({\Varid{step}}^{\Varid{k}}\cdot\Varid{wrap}\cdot\Varid{gstep})\;\Varid{xs}\mathbin{:}\Varid{xs'}\unlhd\Varid{zs}))~~.}
\end{align*}
Proving a proposition containing universal and existential quantification can be thought of as playing a game.
The opponent challenges us by providing \ensuremath{\Varid{xs}} and a way of removing \ensuremath{\mathrm{1}\mathbin{+}\Varid{k}} elements to obtain \ensuremath{\Varid{xs'}}.
We win by presenting a way of removing \ensuremath{\Varid{k}} elements from \ensuremath{\Varid{gstep}\;\Varid{xs}}, such that the result \ensuremath{\Varid{zs}} satisfies \ensuremath{\Varid{xs'}\unlhd\Varid{zs}}.
Equivalently, we present a way of removing \ensuremath{\mathrm{1}\mathbin{+}\Varid{k}} elements from \ensuremath{\Varid{xs}}, while making sure that the hill foot of \ensuremath{\Varid{xs}} is among the \ensuremath{\mathrm{1}\mathbin{+}\Varid{k}} elements removed.
To prove \eqref{eq:step-gstep} is to come up with a strategy to always win the game.

We could just invent the strategy and argue for its correctness.
However, we experimented with another approach: could a proof assistant offer some help?
Can we conjecture the existence of a function that, given the opponent's input, computes \ensuremath{\Varid{zs}}, and try to develop the function and the proof that \ensuremath{\Varid{xs'}\unlhd\Varid{zs}} at the same time, letting their developments mutually guide each other?
It would be a modern realisation of Dijkstra's belief that a program and its proof should be developed hand-in-hand \citep{Dijkstra:74:Programming}.

\paragraph*{The datatypes.}~~
We modeled the problem in the dependently typed language/proof assistant Agda.
For the setting-up, we need to define a number of datatypes.
Firstly, given a type \ensuremath{\Varid{a}} with a total ordering \ensuremath{(\leq )} (which derives a strict ordering \ensuremath{(\mathbin{<})}), lexicographic ordering for \ensuremath{\Conid{List}\;\Varid{a}} is defined by:%
\footnote{
To be consistent with earlier parts of this pearl, we use Haskell-like notations for the Agda code.
Typing relation is denote by \ensuremath{(\mathbin{::})} and list cons by \ensuremath{(\mathbin{:})}.
The two constructors of \ensuremath{\Conid{Nat}} are \ensuremath{\mathrm{0}} and \ensuremath{(\mathrm{1}\mathbin{+})}.
Universally quantified implicit arguments in constructor and function declarations are omitted.
}
\begin{hscode}\SaveRestoreHook
\column{B}{@{}>{\hspre}l<{\hspost}@{}}%
\column{3}{@{}>{\hspre}c<{\hspost}@{}}%
\column{3E}{@{}l@{}}%
\column{11}{@{}>{\hspre}l<{\hspost}@{}}%
\column{41}{@{}>{\hspre}l<{\hspost}@{}}%
\column{E}{@{}>{\hspre}l<{\hspost}@{}}%
\>[B]{}\mathbf{data}\;\_{\unlhd}\_\mathbin{::}\Conid{List}\;\Varid{a}\to \Conid{List}\;\Varid{a}\to \Conid{Set}\;\mathbf{where}{}\<[E]%
\\
\>[B]{}\hsindent{3}{}\<[3]%
\>[3]{}[]_{\unlhd}{}\<[3E]%
\>[11]{}~\mathbin{::}~[\mskip1.5mu \mskip1.5mu]\mathrel{\unlhd}\Varid{xs}{}\<[E]%
\\
\>[B]{}\hsindent{3}{}\<[3]%
\>[3]{}{<}_{\unlhd}{}\<[3E]%
\>[11]{}~\mathbin{::}~\Varid{x}\mathbin{<}\Varid{y}{}\<[41]%
\>[41]{}\to (\Varid{x}\mathbin{:}\Varid{xs})\mathrel{\unlhd}(\Varid{y}\mathbin{:}\Varid{ys}){}\<[E]%
\\
\>[B]{}\hsindent{3}{}\<[3]%
\>[3]{}{\equiv}_{\unlhd}{}\<[3E]%
\>[11]{}~\mathbin{::}~\Varid{xs}\mathrel{\unlhd}\Varid{ys}{}\<[41]%
\>[41]{}\to (\Varid{x}\mathbin{:}\Varid{xs})\mathrel{\unlhd}(\Varid{x}\mathbin{:}\Varid{ys})~~,{}\<[E]%
\ColumnHook
\end{hscode}\resethooks
that is, \ensuremath{[\mskip1.5mu \mskip1.5mu]} is no larger than any lists, \ensuremath{(\Varid{x}\mathbin{:}\Varid{xs})\mathrel{\unlhd}(\Varid{y}\mathbin{:}\Varid{ys})} if \ensuremath{\Varid{x}\mathbin{<}\Varid{y}}, and two lists having the same head is compared by their tails.

Secondly, rather than actually deleting elements of a list, in proofs it helps to remember which elements are deleted. The following datatype \ensuremath{\Conid{Dels}\;\Varid{k}\;\Varid{xs}} can be seen as instructions on how \ensuremath{\Varid{k}} elements are deleted from \ensuremath{\Varid{xs}}:
\begin{hscode}\SaveRestoreHook
\column{B}{@{}>{\hspre}l<{\hspost}@{}}%
\column{3}{@{}>{\hspre}l<{\hspost}@{}}%
\column{9}{@{}>{\hspre}l<{\hspost}@{}}%
\column{E}{@{}>{\hspre}l<{\hspost}@{}}%
\>[B]{}\mathbf{data}\;\Conid{Dels}\mathbin{::}\Conid{Nat}\to \Conid{List}\;\Varid{a}\to \Conid{Set}\;\mathbf{where}{}\<[E]%
\\
\>[B]{}\hsindent{3}{}\<[3]%
\>[3]{}\Varid{end}{}\<[9]%
\>[9]{}\mathbin{::}\Conid{Dels}\;0\;[\mskip1.5mu \mskip1.5mu]{}\<[E]%
\\
\>[B]{}\hsindent{3}{}\<[3]%
\>[3]{}\Varid{keep}{}\<[9]%
\>[9]{}\mathbin{::}\Conid{Dels}\;\Varid{k}\;\Varid{xs}\to \Conid{Dels}\;\Varid{k}\;(\Varid{x}\mathbin{:}\Varid{xs}){}\<[E]%
\\
\>[B]{}\hsindent{3}{}\<[3]%
\>[3]{}\Varid{del}{}\<[9]%
\>[9]{}\mathbin{::}\Conid{Dels}\;\Varid{k}\;\Varid{xs}\to \Conid{Dels}\;(1 + \Varid{k})\;(\Varid{x}\mathbin{:}\Varid{xs})~~.{}\<[E]%
\ColumnHook
\end{hscode}\resethooks
For example, letting \ensuremath{\Varid{xs}\mathrel{=}\text{\ttfamily \char34 abcde\char34}}, \ensuremath{\Varid{ds}\mathrel{=}\Varid{keep}\;(\Varid{del}\;(\Varid{keep}\;(\Varid{del}\;(\Varid{keep}\;\Varid{end}))))\mathbin{::}\Conid{Dels}\;\mathrm{2}\;\Varid{xs}} says that the \ensuremath{\mathrm{1}}st and \ensuremath{\mathrm{3}}rd elements of \ensuremath{\Varid{xs}} (counting from \ensuremath{\mathrm{0}}) are to be deleted.
The function \ensuremath{\Varid{dels}} actually carries out the instruction:
\begin{hscode}\SaveRestoreHook
\column{B}{@{}>{\hspre}l<{\hspost}@{}}%
\column{16}{@{}>{\hspre}l<{\hspost}@{}}%
\column{22}{@{}>{\hspre}l<{\hspost}@{}}%
\column{26}{@{}>{\hspre}l<{\hspost}@{}}%
\column{27}{@{}>{\hspre}l<{\hspost}@{}}%
\column{E}{@{}>{\hspre}l<{\hspost}@{}}%
\>[B]{}\Varid{dels}\mathbin{::}(\Varid{xs}\mathbin{::}\Conid{List}\;\Varid{a})\to \Conid{Dels}\;\Varid{k}\;\Varid{xs}\to \Conid{List}\;\Varid{a}{}\<[E]%
\\
\>[B]{}\Varid{dels}\;[\mskip1.5mu \mskip1.5mu]\;{}\<[16]%
\>[16]{}\Varid{end}{}\<[26]%
\>[26]{}\mathrel{=}[\mskip1.5mu \mskip1.5mu]{}\<[E]%
\\
\>[B]{}\Varid{dels}\;(\Varid{x}\mathbin{:}\Varid{xs})\;{}\<[16]%
\>[16]{}(\Varid{del}\;{}\<[22]%
\>[22]{}\Varid{ds}){}\<[27]%
\>[27]{}\mathrel{=}\Varid{dels}\;\Varid{xs}\;\Varid{ds}{}\<[E]%
\\
\>[B]{}\Varid{dels}\;(\Varid{x}\mathbin{:}\Varid{xs})\;{}\<[16]%
\>[16]{}(\Varid{keep}\;\Varid{ds}){}\<[27]%
\>[27]{}\mathrel{=}\Varid{x}\mathbin{:}\Varid{dels}\;\Varid{xs}\;\Varid{ds}~~.{}\<[E]%
\ColumnHook
\end{hscode}\resethooks
For example, \ensuremath{\Varid{dels}\;\Varid{xs}\;\Varid{ds}\mathrel{=}\text{\ttfamily \char34 ace\char34}}.

Thirdly, the predicate \ensuremath{\Conid{HFoot}\;\Varid{i}\;\Varid{xs}} holds if the \ensuremath{\Varid{i}}-th element in \ensuremath{\Varid{xs}} is the hill foot, that is, the element that would be removed by \ensuremath{\Varid{gstep}\;\Varid{xs}}:
\begin{hscode}\SaveRestoreHook
\column{B}{@{}>{\hspre}l<{\hspost}@{}}%
\column{3}{@{}>{\hspre}l<{\hspost}@{}}%
\column{9}{@{}>{\hspre}l<{\hspost}@{}}%
\column{20}{@{}>{\hspre}l<{\hspost}@{}}%
\column{E}{@{}>{\hspre}l<{\hspost}@{}}%
\>[B]{}\mathbf{data}\;\Conid{HFoot}\mathbin{::}\Conid{Nat}\to \Conid{List}\;\Varid{a}\to \Conid{Set}\;\mathbf{where}{}\<[E]%
\\
\>[B]{}\hsindent{3}{}\<[3]%
\>[3]{}\Varid{last}{}\<[9]%
\>[9]{}\mathbin{::}\Conid{HFoot}\;0\;(\Varid{x}\mathbin{:}[\mskip1.5mu \mskip1.5mu]){}\<[E]%
\\
\>[B]{}\hsindent{3}{}\<[3]%
\>[3]{}\Varid{this}{}\<[9]%
\>[9]{}\mathbin{::}\Varid{x}\mathbin{<}\Varid{y}{}\<[20]%
\>[20]{}\to \Conid{HFoot}\;0\;(\Varid{x}\mathbin{:}\Varid{y}\mathbin{:}\Varid{xs}){}\<[E]%
\\
\>[B]{}\hsindent{3}{}\<[3]%
\>[3]{}\Varid{next}{}\<[9]%
\>[9]{}\mathbin{::}\Varid{x}\geq \Varid{y}{}\<[20]%
\>[20]{}\to \Conid{HFoot}\;\Varid{i}\;(\Varid{y}\mathbin{:}\Varid{xs})\to \Conid{HFoot}\;(1 + \Varid{i})\;(\Varid{x}\mathbin{:}\Varid{y}\mathbin{:}\Varid{xs})~~.{}\<[E]%
\ColumnHook
\end{hscode}\resethooks
For example, \ensuremath{\Varid{next}\;(\Varid{next}\;(\Varid{next}\;(\Varid{next}\;\Varid{this})))} may have type \ensuremath{\Conid{HFoot}\;\mathrm{4}\;\text{\ttfamily \char34 8766678\char34}}, since the \ensuremath{\mathrm{4}}th element is the last in the descending prefix \ensuremath{\text{\ttfamily \char34 87666\char34}}.

Finally, we define a datatype \ensuremath{\Conid{IsDel}\mathbin{::}\Conid{Nat}\to \Conid{Dels}\;\Varid{k}\;\Varid{xs}\to \Conid{Set}} such that, for all \ensuremath{\Varid{ds}\mathbin{:}\Conid{Dels}\;\Varid{k}\;\Varid{xs}}, the relation \ensuremath{\Conid{IsDel}\;\Varid{i}\;\Varid{ds}} holds if \ensuremath{\Varid{ds}} instructs that the \ensuremath{\Varid{i}}-th element of \ensuremath{\Varid{xs}} is to be deleted.
Its definition is repetitive and thus omitted.

\paragraph*{The function and the proofs.}~~
The aim is to construct the following function \ensuremath{\Varid{alter}}:
\begin{hscode}\SaveRestoreHook
\column{B}{@{}>{\hspre}l<{\hspost}@{}}%
\column{E}{@{}>{\hspre}l<{\hspost}@{}}%
\>[B]{}\Varid{alter}\mathbin{::}\Conid{Dels}\;(1 + \Varid{k})\;\Varid{xs}\to \Conid{HFoot}\;\Varid{i}\;\Varid{xs}\to \Conid{Dels}\;(1 + \Varid{k})\;\Varid{xs}~~.{}\<[E]%
\ColumnHook
\end{hscode}\resethooks
It takes an instruction, given by the opponent, that deletes \ensuremath{\mathrm{1}\mathbin{+}\Varid{k}} elements from \ensuremath{\Varid{xs}}, and an evidence that the \ensuremath{\Varid{i}}-th element of \ensuremath{\Varid{xs}} is its hill foot, and produces a possibly altered instruction that also deletes \ensuremath{\mathrm{1}\mathbin{+}\Varid{k}} elements.
Recalling the discussion in the beginning of this section, \ensuremath{\Varid{alter}} should satisfy two properties:
\begin{hscode}\SaveRestoreHook
\column{B}{@{}>{\hspre}l<{\hspost}@{}}%
\column{9}{@{}>{\hspre}l<{\hspost}@{}}%
\column{E}{@{}>{\hspre}l<{\hspost}@{}}%
\>[B]{}\Varid{mono}{}\<[9]%
\>[9]{}\mathbin{::}(\Varid{ds}\mathbin{::}\Conid{Dels}\;(1 + \Varid{k})\;\Varid{xs})\to (\Varid{ft}\mathbin{::}\Conid{HFoot}\;\Varid{i}\;\Varid{xs})\to \Varid{dels}\;\Varid{xs}\;\Varid{ds}\mathrel{\unlhd}\Varid{dels}\;\Varid{xs}\;(\Varid{alter}\;\Varid{ds}\;\Varid{ft})~~,{}\<[E]%
\\
\>[B]{}\Varid{unfoot}{}\<[9]%
\>[9]{}\mathbin{::}(\Varid{ds}\mathbin{::}\Conid{Dels}\;(1 + \Varid{k})\;\Varid{xs})\to (\Varid{ft}\mathbin{::}\Conid{HFoot}\;\Varid{i}\;\Varid{xs})\to \Conid{IsDel}\;\Varid{i}\;(\Varid{alter}\;\Varid{ds}\;\Varid{ft})~~.{}\<[E]%
\ColumnHook
\end{hscode}\resethooks
Given \ensuremath{\Varid{ds}} and \ensuremath{\Varid{ft}}, the property \ensuremath{\Varid{mono}} says that \ensuremath{\Varid{alter}\;\Varid{ds}\;\Varid{ft}} always produces a list that is not worse than that produced by \ensuremath{\Varid{ds}}, while \ensuremath{\Varid{unfoot}} says that \ensuremath{\Varid{alter}\;\Varid{ds}\;\Varid{ft}} does delete the hill foot.

The goal now is to develop \ensuremath{\Varid{alter}}, \ensuremath{\Varid{mono}}, and \ensuremath{\Varid{unfoot}} together.
The reader is invited to give it a try ---
it is more fun trying it yourself!
\footnote{The Agda code can be downloaded from \url{https://scm.iis.sinica.edu.tw/home/2020/dropping-digits/}. }
In most of the steps, the type and proof constraints leave us with only one reasonable choice, while in one case we are led to discover a lemma.
The cases to consider are:%
\footnote{Curly brackets are used in Agda to mention implicit arguments. In each case here we pattern match \ensuremath{\{\mskip1.5mu \Varid{xs}\mathrel{=}\mathbin{...}\mskip1.5mu\}} such that the readers know what input list we are dealing with.}

\begin{enumerate}
\item \ensuremath{\Varid{alter}\;\{\mskip1.5mu \Varid{xs}\mathrel{=}\Varid{x}\mathbin{:}\Varid{y}\mathbin{:}\Varid{ys}\mskip1.5mu\}\;(\Varid{keep}\;\Varid{ds})\;(\Varid{this}\;x{<}y)} --- the opponent keeps  \ensuremath{\Varid{x}}, which is the hill foot because \ensuremath{\Varid{x}\mathbin{<}\Varid{y}}. Due to \ensuremath{\Varid{unfoot}}, we have to delete \ensuremath{\Varid{x}}; a simple way to satisfy \ensuremath{\Varid{mono}} is to keep \ensuremath{\Varid{y}}. Thus we return \ensuremath{\Varid{del}\;(\Varid{keep}\;\Varid{ds'})}, where \ensuremath{\Varid{ds'}} can be any instruction that deletes \ensuremath{\Varid{k}} elements in \ensuremath{\Varid{xs}} --- it doesn't matter how \ensuremath{\Varid{ds'}} does it!
\item \ensuremath{\Varid{alter}\;\{\mskip1.5mu \Varid{xs}\mathrel{=}\Varid{x}\mathbin{:}\Varid{y}\mathbin{:}\Varid{ys}\mskip1.5mu\}\;(\Varid{keep}\;\Varid{ds})\;(\Varid{next}\;x{\geq}y\;\Varid{ft})} --- the opponent keeps \ensuremath{\Varid{x}}, and we have not reached the hill foot yet. In this case it is safe to imitate the opponent and keep \ensuremath{\Varid{x}} too, before recursively calling \ensuremath{\Varid{alter}} to generate the rest of the instruction.
\item \ensuremath{\Varid{alter}\;\{\mskip1.5mu \Varid{xs}\mathrel{=}[\mskip1.5mu \Varid{x}\mskip1.5mu]\mskip1.5mu\}\;(\Varid{del}\;\Varid{end})\;\Varid{last}} --- the opponent deletes the sole element \ensuremath{\Varid{x}}. In this case we delete \ensuremath{\Varid{x}} too, returning \ensuremath{\Varid{del}\;\Varid{end}}.
\item \ensuremath{\Varid{alter}\;\{\mskip1.5mu \Varid{xs}\mathrel{=}\Varid{x}\mathbin{:}\Varid{y}\mathbin{:}\Varid{ys}\mskip1.5mu\}\;(\Varid{del}\;\Varid{ds})\;(\Varid{this}\;x{<}y)} --- the element \ensuremath{\Varid{x}} is the hill foot, and is deleted by the opponent.
In this case, since \ensuremath{\Varid{unfoot}} is satisfied, we can do exactly the same.
We end up returning the same instruction as the opponent's but it is fine, since both \ensuremath{\Varid{mono}} and \ensuremath{\Varid{unfoot}} are satisfied.
\item \ensuremath{\Varid{alter}\;\{\mskip1.5mu \Varid{xs}\mathrel{=}\Varid{x}\mathbin{:}\Varid{y}\mathbin{:}\Varid{ys}\mskip1.5mu\}\;(\Varid{del}\;\Varid{ds})\;(\Varid{next}\;x{\geq}y\;\Varid{ft})} --- the opponent deletes \ensuremath{\Varid{x}}, which is in the descending prefix but is not the hill foot.
This turns out to be the most complex case.
One may try to imitate and delete \ensuremath{\Varid{x}} as well, returning \ensuremath{\Varid{del}\;\Varid{ds'}} for some \ensuremath{\Varid{ds'}}.
However, \ensuremath{\Varid{ds'}}, having type \ensuremath{\Conid{Dels}\;\Varid{k}\;(\Varid{y}\mathbin{:}\Varid{ys})}, cannot be produced by a recursive call to \ensuremath{\Varid{alter}}, whose return type is \ensuremath{\Conid{Dels}\;(1 + \Varid{k})\;\anonymous }.
It could be the case that \ensuremath{\Varid{k}} is \ensuremath{0} and, since we have not deleted the hill foot yet, returning \ensuremath{\Conid{Dels}\;\mathrm{0}\;(\Varid{y}\mathbin{:}\Varid{ys})} would violate \ensuremath{\Varid{unfoot}}.
The lesson learnt from the type is that we can only delete \ensuremath{\mathrm{1}\mathbin{+}\Varid{k}} elements, and we have to save at least one \ensuremath{\Varid{del}} for the hill foot, which is yet to come.
We thus have to further distinguish between two cases:
  \begin{enumerate}
  \item \ensuremath{\Varid{k}\mathrel{=}1 + \Varid{k'}} for some \ensuremath{\Varid{k'}}. In this case we still have room to delete more elements, thus we can safely imitate the opponent, delete \ensuremath{\Varid{x}}, and recursively call \ensuremath{\Varid{alter}}.
  \item \ensuremath{\Varid{k}\mathrel{=}0}. In this case we keep \ensuremath{\Varid{x}}, returning
  \ensuremath{\Varid{keep}\;(\Varid{delfoot}\;\Varid{ft})\mathbin{::}\Conid{Dels}\;\mathrm{1}\;(\Varid{y}\mathbin{:}\Varid{ys})}, where \ensuremath{\Varid{delfoot}\mathbin{::}\Conid{HFoot}\;\Varid{i}\;\Varid{zs}\to \Conid{Dels}\;\mathrm{1}\;\Varid{zs}} computes an instruction that deletes exactly one element, the hill foot.
  What is left to prove to establish \ensuremath{\Varid{mono}} for this case can be extracted to be a lemma:
  \begin{hscode}\SaveRestoreHook
\column{B}{@{}>{\hspre}l<{\hspost}@{}}%
\column{5}{@{}>{\hspre}l<{\hspost}@{}}%
\column{17}{@{}>{\hspre}l<{\hspost}@{}}%
\column{19}{@{}>{\hspre}l<{\hspost}@{}}%
\column{E}{@{}>{\hspre}l<{\hspost}@{}}%
\>[5]{}\Varid{monoAux}\mathbin{::}{}\<[17]%
\>[17]{}\Varid{x}\geq \Varid{y}\to (\Varid{ft}\mathbin{::}\Conid{HFoot}\;\Varid{i}\;(\Varid{y}\mathbin{:}\Varid{ys}))\to {}\<[E]%
\\
\>[17]{}\hsindent{2}{}\<[19]%
\>[19]{}(\Varid{y}\mathbin{:}\Varid{ys})\mathrel{\unlhd}(\Varid{x}\mathbin{:}\Varid{dels}\;(\Varid{y}\mathbin{:}\Varid{ys})\;(\Varid{delfoot}\;\Varid{ft}))~~,{}\<[E]%
\ColumnHook
\end{hscode}\resethooks
  whose proof is an induction on \ensuremath{\Varid{ys}}, keeping \ensuremath{\Varid{x}\geq \Varid{y}} as an invariant.
  If \ensuremath{\Varid{x}\mathbin{<}\Varid{y}} or \ensuremath{\Varid{y}} is the hill foot, we are done.
  Otherwise \ensuremath{\Varid{x}\mathrel{=}\Varid{y}} and we inductively inspect the tail \ensuremath{\Varid{ys}}.
  Without Agda, it would not be easy to discover this lemma.
  \end{enumerate}
\end{enumerate}

\begin{figure}[t]
{\center\small
\begin{hscode}\SaveRestoreHook
\column{B}{@{}>{\hspre}l<{\hspost}@{}}%
\column{18}{@{}>{\hspre}l<{\hspost}@{}}%
\column{34}{@{}>{\hspre}l<{\hspost}@{}}%
\column{43}{@{}>{\hspre}l<{\hspost}@{}}%
\column{E}{@{}>{\hspre}l<{\hspost}@{}}%
\>[B]{}\Varid{alter}\mathbin{::}\Conid{Dels}\;(1 + \Varid{k})\;\Varid{xs}\to \Conid{HFoot}\;\Varid{i}\;\Varid{xs}\to \Conid{Dels}\;(1 + \Varid{k})\;\Varid{xs}{}\<[E]%
\\
\>[B]{}\Varid{alter}\;(\Varid{keep}\;\Varid{ds})\;{}\<[18]%
\>[18]{}(\Varid{this}\;x{<}y){}\<[34]%
\>[34]{}\mathrel{=}\Varid{del}\;(\Varid{keep}\;(\Varid{deleteAny}\;\Varid{ds})){}\<[E]%
\\
\>[B]{}\Varid{alter}\;(\Varid{keep}\;\Varid{ds})\;{}\<[18]%
\>[18]{}(\Varid{next}\;x{\geq}y\;\Varid{ft}){}\<[34]%
\>[34]{}\mathrel{=}\Varid{keep}\;(\Varid{alter}\;\Varid{ds}\;\Varid{ft}){}\<[E]%
\\
\>[B]{}\Varid{alter}\;(\Varid{del}\;\Varid{end})\;{}\<[18]%
\>[18]{}\Varid{last}{}\<[34]%
\>[34]{}\mathrel{=}\Varid{del}\;\Varid{end}{}\<[E]%
\\
\>[B]{}\Varid{alter}\;(\Varid{del}\;\Varid{ds})\;{}\<[18]%
\>[18]{}(\Varid{this}\;x{<}y){}\<[34]%
\>[34]{}\mathrel{=}\Varid{del}\;\Varid{ds}{}\<[E]%
\\
\>[B]{}\Varid{alter}\;\{\mskip1.5mu \Varid{k}\mathrel{=}0\mskip1.5mu\}\;{}\<[18]%
\>[18]{}(\Varid{del}\;\Varid{ds})\;(\Varid{next}\;x{\geq}y\;\Varid{ft}){}\<[43]%
\>[43]{}\mathrel{=}\Varid{keep}\;(\Varid{delfoot}\;\Varid{ft}){}\<[E]%
\\
\>[B]{}\Varid{alter}\;\{\mskip1.5mu \Varid{k}\mathrel{=}1 + \Varid{k}\mskip1.5mu\}\;{}\<[18]%
\>[18]{}(\Varid{del}\;\Varid{ds})\;(\Varid{next}\;x{\geq}y\;\Varid{ft}){}\<[43]%
\>[43]{}\mathrel{=}\Varid{del}\;(\Varid{alter}\;\Varid{ds}\;\Varid{ft})~~,{}\<[E]%
\ColumnHook
\end{hscode}\resethooks
}
\begin{center}
\includegraphics[width=0.8\textwidth]{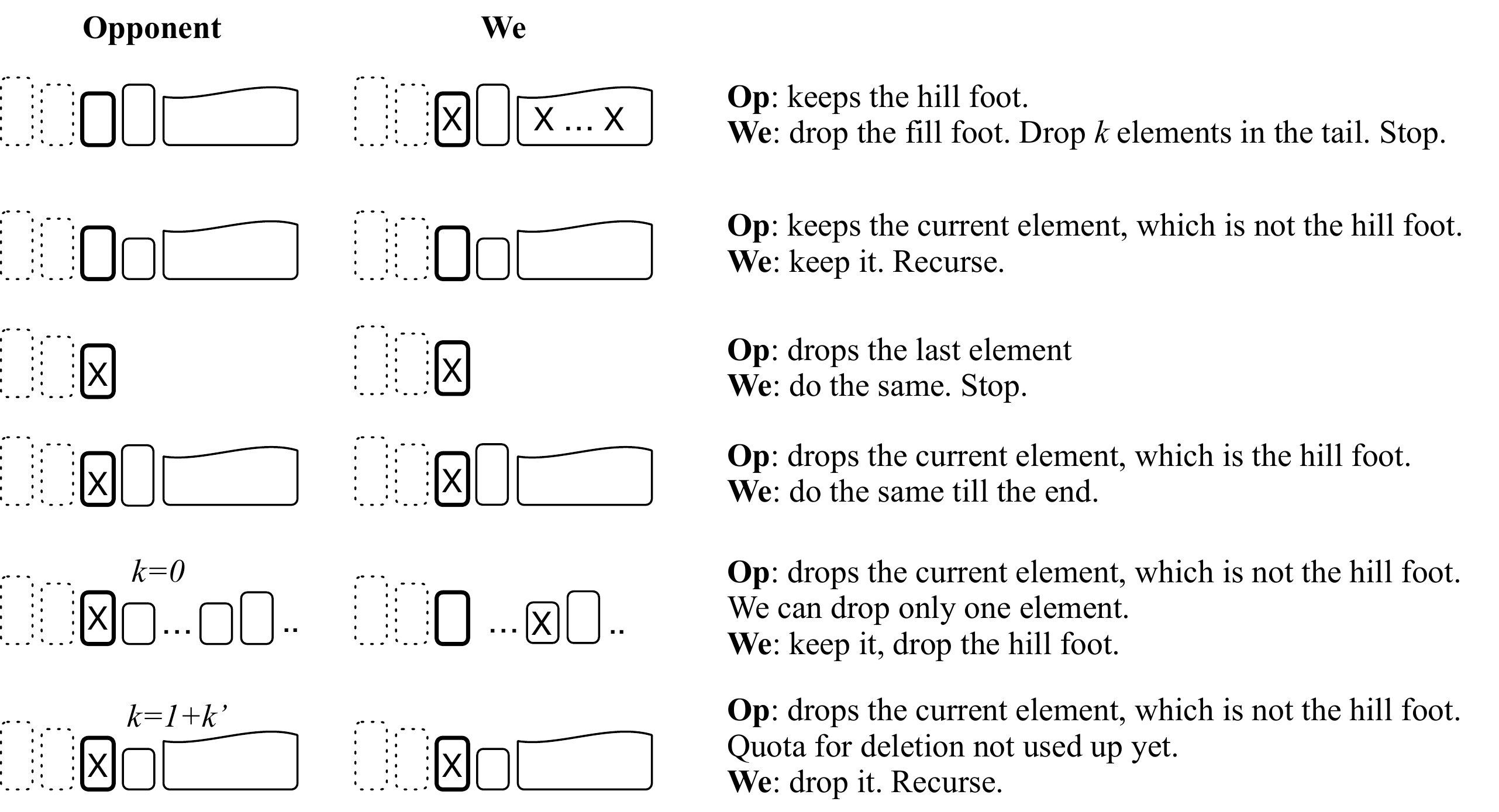}
\end{center}
\caption{The function \ensuremath{\Varid{alter}}, where \ensuremath{\Varid{deleteAny}\;\Varid{ds}} generates a \ensuremath{\Conid{Dels}\;\Varid{k}\;\Varid{xs}}, and a graphical summary.
Element with dotted outline are those that are considered already;
the one with thick outline is the current element.
It is the hill foot if it is smaller than the element to the right, or it is the last.
Deleted elements are marked with a cross.
}
\label{fig:alter}
\end{figure}


In summary, the function \ensuremath{\Varid{alter}} we have constructed, and a graphical summary, are shown in Figure~\ref{fig:alter}.

{\bf Remark}: We may also tuple \ensuremath{\Varid{alter}} and the properties together, and try to construct:
\begin{hscode}\SaveRestoreHook
\column{B}{@{}>{\hspre}l<{\hspost}@{}}%
\column{9}{@{}>{\hspre}l<{\hspost}@{}}%
\column{11}{@{}>{\hspre}l<{\hspost}@{}}%
\column{17}{@{}>{\hspre}l<{\hspost}@{}}%
\column{E}{@{}>{\hspre}l<{\hspost}@{}}%
\>[B]{}\Varid{alter'}{}\<[9]%
\>[9]{}\mathbin{::}(\Varid{ds}\mathbin{::}\Conid{Dels}\;(1 + \Varid{k})\;\Varid{xs})\to (\Varid{ft}\mathbin{::}\Conid{HFoot}\;\Varid{i}\;\Varid{xs})\to {}\<[E]%
\\
\>[9]{}\hsindent{2}{}\<[11]%
\>[11]{}\exists\;(\lambda {}\<[17]%
\>[17]{}(\Varid{ds'}\mathbin{::}\Conid{Dels}\;(1 + \Varid{k})\;\Varid{xs})\to (\Varid{dels}\;\Varid{xs}\;\Varid{ds}\mathrel{\unlhd}\Varid{dels}\;\Varid{xs}\;\Varid{ds'})\times\Conid{IsDel}\;\Varid{i}\;\Varid{ds'})~~.{}\<[E]%
\ColumnHook
\end{hscode}\resethooks
An advantage is that the code of each case of \ensuremath{\Varid{alter}} are next to its proof. A disadvantage is that having to pattern-match the result of \ensuremath{\Varid{alter'}} psychologically discourages one from making a recursive call when needing a \ensuremath{\Conid{Dels}\;(1 + \Varid{k})\;\anonymous }.
It is up to personal preference which style one prefers.

\section{Improving Efficiency}

Back to our code. We have proved that \ensuremath{\Varid{solve}\;\Varid{k}\mathrel{=}{\Varid{gstep}}^{\Varid{k}}}, with \ensuremath{\Varid{gstep}} given by:
\begin{hscode}\SaveRestoreHook
\column{B}{@{}>{\hspre}l<{\hspost}@{}}%
\column{17}{@{}>{\hspre}c<{\hspost}@{}}%
\column{17E}{@{}l@{}}%
\column{20}{@{}>{\hspre}l<{\hspost}@{}}%
\column{E}{@{}>{\hspre}l<{\hspost}@{}}%
\>[B]{}\Varid{gstep}\;[\mskip1.5mu \Varid{x}\mskip1.5mu]{}\<[17]%
\>[17]{}\mathrel{=}{}\<[17E]%
\>[20]{}[\mskip1.5mu \mskip1.5mu]{}\<[E]%
\\
\>[B]{}\Varid{gstep}\;(\Varid{x}\mathbin{:}\Varid{y}\mathbin{:}\Varid{xs}){}\<[17]%
\>[17]{}\mathrel{=}{}\<[17E]%
\>[20]{}\mathbf{if}\;\Varid{x}\mathbin{<}\Varid{y}\;\mathbf{then}\;\Varid{y}\mathbin{:}\Varid{xs}\;\mathbf{else}\;\Varid{x}\mathbin{:}\Varid{gstep}\;(\Varid{y}\mathbin{:}\Varid{xs})~~.{}\<[E]%
\ColumnHook
\end{hscode}\resethooks
Each time \ensuremath{\Varid{gstep}} is called, it takes \ensuremath{\Conid{O}\;(\Varid{n})} steps to go through the descending prefix and find the hill foot, before the next invocation of \ensuremath{\Varid{gstep}} starts from the beginning of the list again. Therefore, \ensuremath{\Varid{solve}\;\Varid{k}} takes \ensuremath{\Conid{O}\;(\Varid{kn})} steps over all. This is certainly not necessary --- to find the next hill foot, the next \ensuremath{\Varid{gstep}} could start from where the previous one left off.

The way to implement this idea is to bring in an accumulating parameter. Suppose we
generalise \ensuremath{\Varid{solve}} to a function \ensuremath{\Varid{gsolve}}, defined by
\begin{hscode}\SaveRestoreHook
\column{B}{@{}>{\hspre}l<{\hspost}@{}}%
\column{3}{@{}>{\hspre}l<{\hspost}@{}}%
\column{E}{@{}>{\hspre}l<{\hspost}@{}}%
\>[3]{}\Varid{gsolve}\;\Varid{k}\;\Varid{xs}\;\Varid{ys}\mathrel{=}\Varid{solve}\;\Varid{k}\;(\Varid{xs}\mathbin{{+}\mskip-8mu{+}}\Varid{ys})~~,{}\<[E]%
\ColumnHook
\end{hscode}\resethooks
with the proviso that the argument \ensuremath{\Varid{xs}} is constrained to be a descending sequence.
In particular, \ensuremath{\Varid{solve}\;\Varid{k}\;\Varid{xs}\mathrel{=}\Varid{gsolve}\;\Varid{k}\;[\mskip1.5mu \mskip1.5mu]\;\Varid{xs}}.
We aim to develop a recursive definition of \ensuremath{\Varid{gsolve}}. Clearly,
\begin{hscode}\SaveRestoreHook
\column{B}{@{}>{\hspre}l<{\hspost}@{}}%
\column{3}{@{}>{\hspre}l<{\hspost}@{}}%
\column{E}{@{}>{\hspre}l<{\hspost}@{}}%
\>[3]{}\Varid{gsolve}\;\mathrm{0}\;\Varid{xs}\;\Varid{ys}\mathrel{=}\Varid{xs}\mathbin{{+}\mskip-8mu{+}}\Varid{ys}~~.{}\<[E]%
\ColumnHook
\end{hscode}\resethooks
Recalling that \ensuremath{\Varid{gstep}} drops the last element of a descending list, we know that \ensuremath{\Varid{k}} repetitions of \ensuremath{\Varid{gstep}} on a decreasing list will drop the last \ensuremath{\Varid{k}} elements.
Hence
\begin{hscode}\SaveRestoreHook
\column{B}{@{}>{\hspre}l<{\hspost}@{}}%
\column{3}{@{}>{\hspre}l<{\hspost}@{}}%
\column{E}{@{}>{\hspre}l<{\hspost}@{}}%
\>[3]{}\Varid{gsolve}\;\Varid{k}\;\Varid{xs}\;[\mskip1.5mu \mskip1.5mu]\mathrel{=}\Varid{dropLast}\;\Varid{k}\;\Varid{xs}~~,{}\<[E]%
\ColumnHook
\end{hscode}\resethooks
where \ensuremath{\Varid{dropLast}\;\Varid{k}} drops the last \ensuremath{\Varid{k}} elements of a list. We will not give a formal definition
of \ensuremath{\Varid{dropLast}} as it will be replaced by another function in a moment. That deals with the
two base cases. For the recursive case, it is easy to prove the following property of \ensuremath{\Varid{gstep}}:
\begin{hscode}\SaveRestoreHook
\column{B}{@{}>{\hspre}l<{\hspost}@{}}%
\column{3}{@{}>{\hspre}l<{\hspost}@{}}%
\column{23}{@{}>{\hspre}l<{\hspost}@{}}%
\column{50}{@{}>{\hspre}c<{\hspost}@{}}%
\column{50E}{@{}l@{}}%
\column{53}{@{}>{\hspre}c<{\hspost}@{}}%
\column{53E}{@{}l@{}}%
\column{56}{@{}>{\hspre}l<{\hspost}@{}}%
\column{E}{@{}>{\hspre}l<{\hspost}@{}}%
\>[3]{}\Varid{gstep}\;(\Varid{xs}\mathbin{{+}\mskip-8mu{+}}\Varid{y}\mathbin{:}\Varid{ys}){}\<[23]%
\>[23]{}\mid \Varid{null}\;\Varid{xs}\mathrel{\vee}\Varid{last}\;\Varid{xs}\geq \Varid{y}{}\<[50]%
\>[50]{}\enspace{}\<[50E]%
\>[53]{}\mathrel{=}{}\<[53E]%
\>[56]{}\Varid{gstep}\;((\Varid{xs}\mathbin{{+}\mskip-8mu{+}}[\mskip1.5mu \Varid{y}\mskip1.5mu])\mathbin{{+}\mskip-8mu{+}}\Varid{ys}){}\<[E]%
\\
\>[23]{}\mid \Varid{otherwise}{}\<[53]%
\>[53]{}\mathrel{=}{}\<[53E]%
\>[56]{}\Varid{init}\;\Varid{xs}\mathbin{{+}\mskip-8mu{+}}\Varid{y}\mathbin{:}\Varid{ys}~~,{}\<[E]%
\ColumnHook
\end{hscode}\resethooks
which can be used to construct the following case of \ensuremath{\Varid{gsolve}}:
\begin{hscode}\SaveRestoreHook
\column{B}{@{}>{\hspre}l<{\hspost}@{}}%
\column{3}{@{}>{\hspre}l<{\hspost}@{}}%
\column{27}{@{}>{\hspre}l<{\hspost}@{}}%
\column{54}{@{}>{\hspre}c<{\hspost}@{}}%
\column{54E}{@{}l@{}}%
\column{57}{@{}>{\hspre}l<{\hspost}@{}}%
\column{E}{@{}>{\hspre}l<{\hspost}@{}}%
\>[3]{}\Varid{gsolve}\;(\mathrm{1}\mathbin{+}\Varid{k})\;\Varid{xs}\;(\Varid{y}\mathbin{:}\Varid{ys}){}\<[27]%
\>[27]{}\mid \Varid{null}\;\Varid{xs}\mathrel{\vee}\Varid{last}\;\Varid{xs}\geq \Varid{y}{}\<[54]%
\>[54]{}\enspace{}\<[54E]%
\>[57]{}\mathrel{=}\Varid{gsolve}\;(\mathrm{1}\mathbin{+}\Varid{k})\;(\Varid{xs}\mathbin{{+}\mskip-8mu{+}}[\mskip1.5mu \Varid{y}\mskip1.5mu])\;\Varid{ys}{}\<[E]%
\\
\>[27]{}\mid \Varid{otherwise}{}\<[57]%
\>[57]{}\mathrel{=}\Varid{gsolve}\;\Varid{k}\;(\Varid{init}\;\Varid{xs})\;(\Varid{y}\mathbin{:}\Varid{ys})~~.{}\<[E]%
\ColumnHook
\end{hscode}\resethooks

The second optimisation is simply to replace the list \ensuremath{\Varid{xs}} in the definition of \ensuremath{\Varid{gsolve}} by \ensuremath{\Varid{reverse}\;\Varid{xs}} to avoid adding elements at the end of a list.
That leads to our final algorithm:
\begin{hscode}\SaveRestoreHook
\column{B}{@{}>{\hspre}l<{\hspost}@{}}%
\column{21}{@{}>{\hspre}l<{\hspost}@{}}%
\column{24}{@{}>{\hspre}l<{\hspost}@{}}%
\column{49}{@{}>{\hspre}c<{\hspost}@{}}%
\column{49E}{@{}l@{}}%
\column{52}{@{}>{\hspre}l<{\hspost}@{}}%
\column{E}{@{}>{\hspre}l<{\hspost}@{}}%
\>[B]{}\Varid{solve}\;\Varid{k}\;\Varid{xs}\mathrel{=}\Varid{gsolve}\;\Varid{k}\;[\mskip1.5mu \mskip1.5mu]\;\Varid{xs}~~,{}\<[E]%
\\[\blanklineskip]%
\>[B]{}\Varid{gsolve}\;\mathrm{0}\;\Varid{xs}\;\Varid{ys}{}\<[21]%
\>[21]{}\mathrel{=}{}\<[24]%
\>[24]{}\Varid{reverse}\;\Varid{xs}\mathbin{{+}\mskip-8mu{+}}\Varid{ys}{}\<[E]%
\\
\>[B]{}\Varid{gsolve}\;\Varid{k}\;\Varid{xs}\;[\mskip1.5mu \mskip1.5mu]{}\<[21]%
\>[21]{}\mathrel{=}{}\<[24]%
\>[24]{}\Varid{reverse}\;(\Varid{drop}\;\Varid{k}\;\Varid{xs}){}\<[E]%
\\
\>[B]{}\Varid{gsolve}\;\Varid{k}\;\Varid{xs}\;(\Varid{y}\mathbin{:}\Varid{ys}){}\<[21]%
\>[21]{}\mid \Varid{null}\;\Varid{xs}\mathrel{\vee}\Varid{head}\;\Varid{xs}\geq \Varid{y}{}\<[49]%
\>[49]{}\mathrel{=}{}\<[49E]%
\>[52]{}\Varid{gsolve}\;\Varid{k}\;(\Varid{y}\mathbin{:}\Varid{xs})\;\Varid{ys}{}\<[E]%
\\
\>[21]{}\mid \Varid{otherwise}{}\<[49]%
\>[49]{}\mathrel{=}{}\<[49E]%
\>[52]{}\Varid{gsolve}\;(\Varid{k}\mathbin{-}\mathrm{1})\;(\Varid{tail}\;\Varid{xs})\;(\Varid{y}\mathbin{:}\Varid{ys})~~,{}\<[E]%
\ColumnHook
\end{hscode}\resethooks
where \ensuremath{\Varid{drop}\;\Varid{k}} is a standard Haskell function that drops the first \ensuremath{\Varid{k}} elements from a list.
For an operational explanation, \ensuremath{\Varid{gsolve}} traverses through the list, keeping looking for the next hill foot to delete ---
the \ensuremath{\Varid{otherwise}} case is when a hill foot is found.
The list \ensuremath{\Varid{xs}} is the traversed part --- \ensuremath{(\Varid{xs},\Varid{ys})} forms a zipper.
The head of \ensuremath{\Varid{xs}} is a possible candidate of the hill foot.
While the algorithm looks simple once understood, without calculation it is not easy to get the details right.
The authors have come up with several versions that are wrong, before sitting down to calculate it!

To time the program, note that at each step either \ensuremath{\Varid{k}} is reduced to \ensuremath{\Varid{k}\mathbin{-}\mathrm{1}} or \ensuremath{\Varid{y}\mathbin{:}\Varid{ys}} is reduced to \ensuremath{\Varid{ys}}.
Hence \ensuremath{\Varid{solve}\;\Varid{k}\;\Varid{xs}} takes \ensuremath{\Conid{O}\;(\Varid{k}\mathbin{+}\Varid{n})\mathrel{=}\Conid{O}\;(\Varid{n})} steps, where \ensuremath{\Varid{n}\mathrel{=}\Varid{length}\;\Varid{xs}}.


\section{Conclusion}

To construct a linear-time algorithm for solving the puzzle,
various techniques were employed. The structure of the greedy algorithm was proved using predicate logic, and the proof was simplified from relational program calculus. Agda was used to give a constructive proof of the greedy condition, and equational reasoning was used to derive the greedy step as well as the final, linear-time optimisation.

\bibliographystyle{abbrvnat}

\end{document}